# Machine Learning Approach for Detection of nonTor Traffic

Elike Hodo
University of Strathclyde
Scotland
elike.hodo@strath.ac.uk

Xavier Bellekens
University of Abertay Dundee
Scotland
x.bellekens@abertay.ac.uk

Ephraim Iorkyase
University of Strathclyde
Scotland
Ephraim.iorkyase@strath.ac.uk

Andrew Hamilton
University of Strathclyde
Scotland
Andrew.w.hamilton@strath.ac.uk

Christos Tachtatzis
University of Strathclyde
Scotland
Christos.tachtatzis@strath.ac.uk

Robert Atkinson
University of Strathclyde
Scotland
Robert.atkinson@strath.ac.uk

## ABSTRACT

Intrusion detection has attracted a considerable interest from researchers and industries. After many years of research the community still faces the problem of building reliable and efficient intrusion detection systems (IDS) capable of handling large quantities of data with changing patterns in real time situations. The Tor network is popular in providing privacy and security to end user by anonymising the identity of internet users connecting through a series of tunnels and nodes. This work focuses on the classification of Tor traffic and nonTor traffic to expose the activities within Tor traffic that minimizes the protection of users. A study to compare the reliability and efficiency of Artificial Neural Network and Support vector machine in detecting nonTor traffic in UNB-CIC Tor Network Traffic dataset is presented in this paper. The results are analysed based on the overall accuracy, detection rate and false positive rate of the two algorithms. Experimental results show that both algorithms could detect nonTor traffic in the dataset. A hybrid Artificial neural network proved a better classifier than SVM in detecting nonTor traffic in UNB-CIC Tor Network Traffic dataset.

## KEYWORDS

Artificial neural network, support vector machines, intrusion detection systems, Tor and nonTor, UNB-CIC Tor Network Traffic dataset.

## 1  INTRODUCTION

The computing world has changed over the past decade due to the rapid development of internet and new privacy enhancement technologies to circumvent internet censorship. Tor which is popular in fighting internet censorship has been deployed to serve thousands of users transferring terabytes of data daily [1], [2].

Tor is an overlay network designed to provide privacy and anonymity over the internet for TCP based applications like browsing. It operates by anonymising the identity of users connecting through a series of tunnels and nodes. A user browses the web firstly by sending a request anonymously to Tor routers from one of the directory servers [2], [3]. Once a connection is established, traffic is relayed to the first router also called the Entry Guard. A session key is then generated between the client and the Entry Guard using Deffie-Hellman key exchange [4]. The same process is repeated on one router (hop) at a time to extend the circuit each time with established session keys for the previous routers. The last hop called the exit router communicates direct with the destination as a proxy [1]. With the establishment of three routers, the circuit is ready for internet traffic. The design of TOR network which includes the use of three hops and session keys helps to maintain anonymity through a concept called 'perfect forward secrecy'[2], [5].

Tor networks are created to give internet users their privacy, freedom of speech, illegal tapping traffic and surveillance of network threatening users' personal identity [6]. Besides Tor network being used for good, greater portion of its traffic are port scans, hacking attempts, exfiltration of stolen data and online criminality [2].

Over the last decade, Tor traffic classification has advanced in its applications in systems like quality of service (QoS) tools or Security information and Event management (SIEM) [7]. A considerable interest have been attracted from researchers and the industries to the study of these technologies and developing classification techniques [8][7].

To this effect intrusion detection system (IDS) plays an important role in Tor networks. Intrusion Detection Systems are placed on the networks to monitor and detect anomalies. In general IDS can be categorised into two components, based on the detection technique. Signature based and Outlier based IDS. Most IDs employ a signature based detection approach where the network traffic is monitored and compared against database rules or signature of known anomaly in network traffic [9][10]. An alarm is raised on detection of a mismatch. Signature based is the most common as they do not necessarily have to learn the network traffic's behaviour. Although it is effective in detecting known anomalies, it cannot detect unknown anomalies unless the signature and rules are updated with new signatures [11][12]. Signature based is known to have a significant time lapse between anomaly detection and activation of its corresponding signature [10]. Signature based techniques are mainly human-dependent in creating, testing and deploying signatures.

The outlier technique is a behavioural based detection system. It observes changes in normal activity of network traffic and builds a profile of the network traffic being monitored [13][14]. An alarm is raised whenever a deviation from the normal behaviour is detected. It has the ability to detect unknown anomalies. However outlier detection based IDS have the disadvantage of being computational expensive because the profile generated over a period needs to be updated against each system activity [15] [10]. Machine learning techniques have the ability to learn the normal and anomalous patterns automatically by training a dataset to predict an anomaly in network traffic. One important characteristic defining the effectiveness of machine learning techniques is the features extracted from raw data for classification and detection. Features are the important information extracted from raw data. The underlying factor in selecting the best features lies in a trade-off between detection accuracy and false alarm rates. The use of all features on the other hand will lead to a significant overhead and thus reducing the risk of removing important features. Although the importance of feature selection cannot be overlooked, intuitive understanding of the problem is mostly used in the selection of features [16].

This paper analyses the performance of Artificial neural network (ANN) and Support vector machines (SVM) in terms of overall accuracy in detecting nonTor traffic in a Tor network traffic dataset data from the University of New Brunswick (UNB), Canadian Institute for cyber security (CIC) using a hybrid anomaly based approach. As part of the work, the results are compared with the results of A.Lashkari et al.[7] being the only study published to the best of our knowledge using the UNB-CIC Tor Network Traffic dataset [17]. A. Lashkari et al.[7]extracted 23 time based features from the dataset. A combination algorithm Cfs-SubsetEval + BestFirst (SE+BF) and Infogain+Ranker (IG+RK) was used to reduce the number of features from 23 to 5. The results from the feature selection algorithm was used to test different machine learning algorithms (ZeroR, C4.5 and KNN) using 10 fold cross validation and measured the weighted average precision and recall. Their results showed C4.5 was the best classifier.

In the proposed approach 10 features are selected out of the 28 features of the dataset using Correlation based feature selection (CFS) for training and testing the detection algorithms.

The rest of the paper is organised as follows: section 2 describes intrusion detection systems, section 3 describes the UNB-CIC Tor Network Traffic dataset, section 4 introduces Artificial neural network and Support vector machines algorithms used in the experiment respectively, section 5 analysis experimental results, conclusion and future works are presented in section VI.

## 2 INTRUSION DETECTION SYSTEM

Intrusion detection system is a software application or a device placed at strategic places on a network to monitor and detect anomalies in network traffic [1][2] as shown in Fig. 1. The main features of IDS are to raise an alarm when an anomaly is detected. A complementary approach is to take corrective measures when anomalies are detected, such an approach is referred to as an intrusion Prevention System (IPS) [3]. Based on the interactivity property of IDS, it can be designed to work either on-line or off-line. On-line IDS operates on a network in real time by analysing traffic packets and applying rules to classify normal and analogous traffic. Off-line IDS operates by storing data and after processing to classify normal and anomaly.

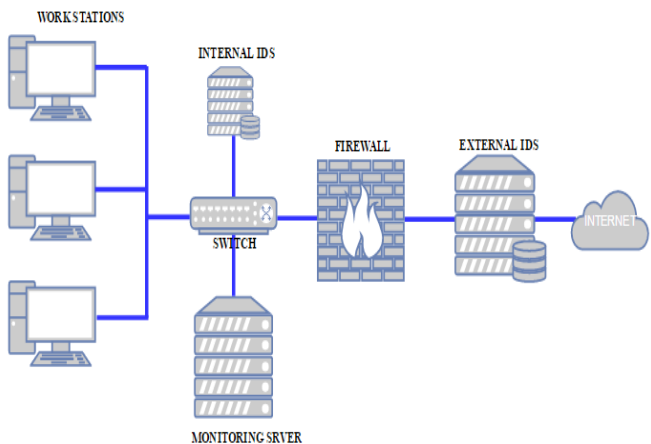

**Figure 1. Intrusion Detection System model.**

## 3 UNB-CIC TOR NETWORK TRAFFIC DATASET

UNB-CIC Tor Network Traffic dataset [17] is a representative dataset of real-world traffic defined as a set of task. Three users were set up for browser traffic collection and two users for the communication parts such as chat, mail, p2p etc. from more than 18 representative applications such as Facebook, skype, Spotify, Gmail etc. The dataset contains 8 types of Tor traffic as shown in table 1 and non-Tor traffic.

### 3.1 UNB-CIC Tor network Traffic dataset description

**Table 1: Description of UNB-CIC Tor Network Traffic**

| Type of Traffic | Description |
|---|---|

| | | Feature Name | Feature description |
|---|---|---|---|
| Browsing | HTTP and HTTPS traffic generated by users while using Firefox and chrome | Source IP | IP address sending packets to destination |
| | | Source Port | Port sending packets from source |
| | | Destination IP | IP address receiving packets from source |
| | | Destination Port | Port receiving packets |
| Email | Traffic samples generated using a Thunderbird client and two other accounts holders. Mails were delivered through SMTP/S and received using POP3/SSL in client 1 and IMAP/SSL in client 2. | Protocol | Type of the protocol used |
| | | Flow Duration | Length of connection in seconds |
| | | Flow Bytes/s | Number of data bytes |
| | | Flow Packets/s | Number of data packets |
| | | Flow IAT Mean | Packets flow inter arrival time Mean |
| | | Flow IAT Std | Packets flow inter arrival time Standard deviation |
| | | Flow IAT Max | Packets flow inter arrival time Max. |
| | | Flow IAT Min | Packets flow inter arrival time Min. |
| Chat | Instant messaging applications were identified under the chat label. The label was associated with Facebook and hangouts through web browser, skype and IAM and ICQ using an application called pidgin. | Fwd IAT Mean | Forward inter arrival time, the time between two packets Sent forward direction Mean. |
| | | Fwd IAT Std | Forward inter arrival time, the time between two packets sent forward direction Standard deviation. |
| | | Fwd IAT Max | Forward inter arrival time, the time between two packets sent forward direction Max. |
| Audio-Streaming | Traffic was captured from Spotify identifying audio applications that require a continuous and steady stream of data. | Fwd IAT Min | Forward inter arrival time, the time between two packets sent forward direction Min. |
| | | Bwd IAT Mean | Backward inter arrival time, the time between two packets sent backward Mean. |
| Video-Streaming | Traffic was captured from YouTube and Vimeo services using Chrome and Firefox identifying video applications that require a continuous and steady stream of data. | Bwd IAT Std | Backward inter arrival time, the time between two packets sent backward Standard deviation. |
| | | Bwd IAT Max | Backward inter arrival time, the time between two packets sent backward Max. |
| | | Bwd IAT Min | Backward inter arrival time, the time between two packets sent backward Min. |
| File Transfer | This traffic was generated from skype file transfers, FTP over SSH (SFTP) and FTP over SSL (FTPS) traffic sessions identifying the traffic applications sending or receiving file documents. | Active Mean | The amount of time a flow was active before becoming idle mean. |
| | | Active Std | The amount of time a flow was active before becoming idle Standard deviation. |
| | | Active Max | The amount of time a flow was active before becoming idle Max. |
| Voice over Internet Protocol (Voip | This is the traffic generated by voice applications using Facebook, Hangouts and Skype. | Active Min | The amount of time a flow was active before becoming idle Min. |
| | | Idle Mean | The amount of time a flow was idle before becoming active Mean. |
| | | Idle Std | The amount of time a flow was idle before becoming active Std deviation. |
| | | Idle Max | The amount of time a flow was idle before becoming active Max. |
| | | Idle Min | The amount of time a flow was idle before becoming active Min. |

**Table 2: Description of captured features**

The non-Tor traffic captured in the dataset contains unique characteristics differentiating it from the Tor traffic. These characteristics are called features. The UNB-CIC Tor Network Traffic dataset contains a total of 28 features listed in table 2.

The features were generated by a sequence of packets having the same values for {source IP, source Port, destination port and protocol (TCP and UDP)}. All Tor traffic was TCP since the flow does not support UDP. The generation of flows was done by a new application, the ISCX Flow Meter which generates bidirectional flows [7].

## 4 DETECTION ALGORITHMS

### 4.1 Artificial Neural Network

Artificial neural network (ANN) consists of information processing elements known to mimic neurons of the brain.

In this experiment, the neural network which is a Multilayer perceptron (MLP) is provided with a set labelled training set which learns a mapping from input features listed in table II represented as $x$ in Fig. 2 to outputs (Tor and NonTor) as $y$ in Fig. 2 given a labelled set of inputs-output pairs

$d = \{(x_i, y_i)\}_{i=1}^{N}$
(1)

Where, $d$ is called the training set and $N$ is the number of training examples. It is assumed that $y_i$ is a categorical variable from some infinite set, $y_i \in \{1 \ldots C\}$ [4]. The technique used to train the MLP neural network is the Back Propagation hence the name MLP-BP.

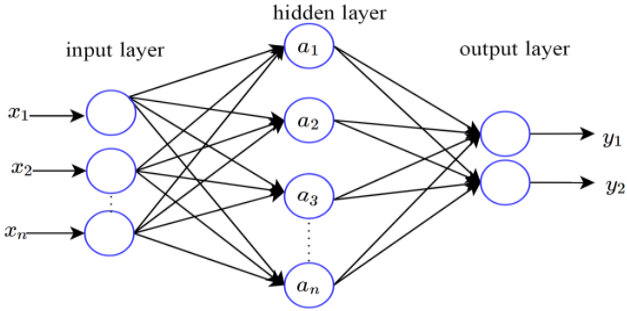

**Figure 2: Artificial Neural Network Model used in experiment.**

The construction of the MLP-BP neural network is by putting layers of non-linear elements to form complex hypotheses. Each node takes an element of a feature vector. The structure of the ANN consists of three layers feed-forward neural network as shown in Fig. 2 Nodes labelled $x_1 \ldots \ldots x_n$ have been used to represent the input feature vectors to the ANN.

Hidden inner nodes $a_1 \ldots \ldots a_n$ make up the hidden layer with an output layer of $y_1$ and $y_2$ nodes denoting different output classes (Tor and NonTor). The interconnection between the nodes is associated with scalar weights with an initial weight assigned to the connection. During training, the weights are adjusted.

Evaluating the hypotheses is done by setting the input modes in a feed-back process and the values are propagated through the network to the output. At this stage the gradient descent is used so as to push the error in the output node back through the network by a back propagation process in order to estimate the error in the hidden nodes. The gradient of the cost – function is then calculated [5].

### 4.2 Support Vector Machines

Support Vector Machines (SVM) is a machine learning algorithm that learns to classify data using points labelled training examples falling into one or two classes. The SVM algorithm builds a model that can predict if a new example falls into one category or the other [6]. The model is constructed by constructing $k$ Models of SVM, where $k$ denotes the number of classes (Tor and NonTor). $x_1$ and $y_2$ SVM represented as $l$th SVM is trained with all the examples in the $l$th class labelled 1 and the other labelled 0. Where, $x_i \in R^d$, $y_i \in \{1,0\}$, $i = 1 \ldots N$ and $y_i \in \{1 \ldots k\}$ is a class of $x_i$. Introducing a slack of positive variables $\xi_i$, that measures the extent of constraint in a non-linear situation. The prima Optimisation problem becomes [7]:

$$\min_{w^l, b^l, \xi^l} \frac{1}{2}(w^l)^T w^l + C \sum_{i=1}^{N} \xi_i^l$$

$(w^l)^T \phi(x)_i + b^l \geq 1 - \xi_i^l, \text{ if } y_i = N,$
(2)
$(w^l)^T \phi(x)_i + b^l \leq -1 + \xi_i^l, \text{ if } y_i = N,$
$\xi_i^l \geq 0, \; i = 1 \ldots \ldots N,$

Where the training set $x_i$ are mapped into higher dimensional space by the function $\phi$ and $C$, where $C$ is a parameter which trades off wide margin with small number of margin failures.

Minimisation of $\frac{1}{2}(w^l)^T w^l$ implies maximising $\frac{2}{\|w^l\|}$, which is the margin between the two data points. The SVM then searches for a balance between the regularisation term $\frac{1}{2}(w^l)^T w^l$ and the errors in training the dataset. Solving (2) gives $k$ decision functions:

$$\begin{matrix} (w)^{1^T} \phi(x) + b^1 \\ \vdots \\ (w)^{k^T} \phi(x) + b^k \end{matrix}$$
(3)

where $x$ is the class having the largest value of the decision function:

$x \equiv argmax_{l \equiv 1 \ldots k}((w^l)^T \phi(x) + b^l)$
(4)

The dual problem of (2) having the same number of variables as the number of data in (2). Thus $k\,N$-variable quadratic programing problems are solved.

Machine Learning Approach for Detection of nonTor Traffic     ARES'17, August 2017, Regio Callabria, ITALY

## 5 EXPERIMENTAL RESULTS ANALYSIS

### 5.1 Results Evaluation Metrics

The effectiveness of IDS requires high accuracy, high detection rate (Recall) and high Positive Predictive value (Precision) as well as low false positive rate. The performance of IDS in general is evaluated in terms of overall accuracy, detection rate and false positive rate.

Accuracy (ACC) = $\frac{TP}{TP+TN+FP+FN}$

Detection Rate (DR) = $\frac{TP}{TP+FP}$

False Positive rate (FPR) = $\frac{FP}{FP+TN}$

Positive Predictive Value (PPV) = $\frac{TP}{TP+FP}$

Where, True Negative (TN): a measure of the number of normal events rightly classified normal.

True Positive (TP): a measure of attacks classified rightly as attack.

False Positive (FP): a measure of normal events misclassified as attacks.

False Negative (FN): a measure of attacks misclassified as normal.

### 5.2 Feature Selection Algorithm

This paper proposes correlation based feature selection (CFS) to select the relevant features out of the 28 features.

CFS is a filtering algorithm using a correlation based heuristic evaluation function to rank feature subsets. A good set of features are highly correlated with the class (target) and at the same time uncorrelated to each other. Redundant features are ignored because they have low correlation with class and will turn to highly correlated with one or more of the remaining features. A feature is accepted based on the extent it predicts classes in areas of the instance space which has not been predicted by other features.

Equation (8) shows the CFS feature subset evaluation function.

$$M_s = \frac{k\overline{r_{cf}}}{\sqrt{k+k(k-1)\overline{r_{ff}}}} \qquad (5)$$

Where $M_s$ is heuristic "merit" of a feature subset $s$ containing $k$ features, $\overline{r_{cf}}$ is the average feature class correlation. The numerator can be thought of as giving an indication of how predictive a group of features are; the denominator of how much redundancy there is among the features [8].

The CFS algorithm reduces the dimensionality of the dataset, reduces overfitting and gives a shorter training time. Table IV shows the 10 selected features based on the appropriate correlation measure and heuristic search strategy. The selected features are: destination Port, Bwd IAT Mean, Idle Max., Fwd IAT Min., Source Port, idel Min., Flow Bytes/s, Flow IAT Std., Source IP, Destination IP.

### 5.3 Experimental Results

Neural network and Support vector machine classification involves two phases: the classification phase and training phase as shown in Fig. 3. In the training phase, the algorithm learns the distribution of the features with corresponding classes. During the classification phase, the learned model is applied to a test set which has not been previously seen by the training phase.

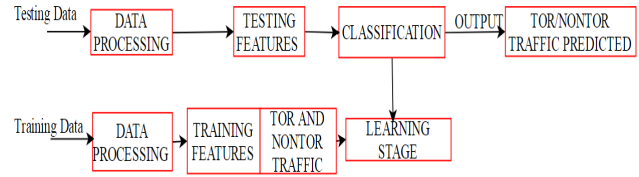

**Figure 3: Experimental Model.**

In this work, experiment was performed by training ANN and SVM with UNB-CIC Tor Network Traffic dataset to detect nonTor Traffic.

In the first set of experiment, ANN was trained with all 28 features of the dataset with 20-hidden neurons and with 10 features selected using CFS with 6-hidden nodes. The ANN uses Levenberg-Marquardt training function (trainlm) for learning.

In the second set of the experiment SVM was trained with all 28 features in the dataset and with 10 features selected using CFS.

The performance of ANN and SVM was evaluated on train (70%) dataset, test (15%) dataset and validation 20% dataset.

Fig. 4 shows the results after training ANN and SVM with all 28 features and 10 features selected by CFS. A.Lashkari *et al*.[9] Proposed C 4.5 in classifying Tor Traffic and nonTor Traffic using only the time based features of the dataset.

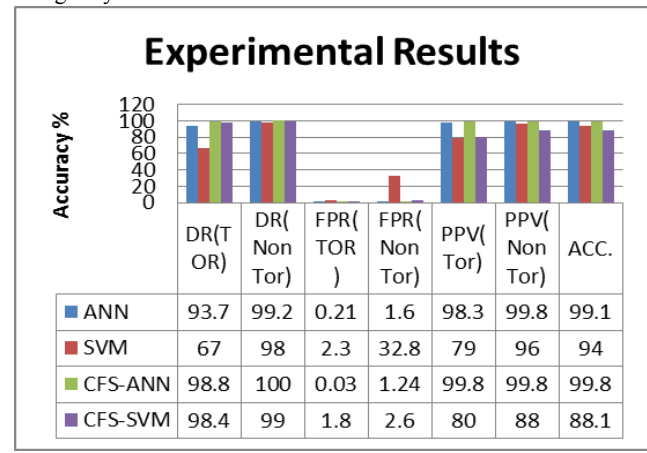

**Figure 4: Experimental Results of ANN, SVM, CFS-ANN and CFS-SVM. .**

The experimental results as compared to the results in [7] show CFS-ANN performs with an overall accuracy of 99.8% in the classification of Tor and nonTor using only 10 features in the dataset. On the other hand the DR for Tor and nonTor Traffic in CFS-ANN recorded 98.9% and 100% respectively which performed better than C 4.5. The best values in detection accuracy, detection rate with a low false positive rate in the classification of Tor and nonTor traffic were recorded by CFS-ANN making it a promising detection system for nonTor traffic. A comparison of experimental results with results from [7] are shown in table 3.

**Table 3: Experimental results compared to A.Lashkari et al.[7]**

| PERFORMANCE | ANN | CFS-ANN | SVM | CFS-SVM | C4.5 [7] |
|---|---|---|---|---|---|
| **DR (Tor) %** | 93.7 | 98.8 | 67 | 98.4 | 93.4 |
| **FPR (Tor) %** | 0.21 | 0.03 | 2.3 | 1.8 | - |
| **PPV (Tor) %** | 98.3 | 99.8 | 79 | 80 | 94.8 |
| **DR (nonTor) %** | 99.2 | 100 | 98 | 99 | 99.2 |
| **FPR (nonTor) %** | 1.6 | 1.2 | 32.8 | 2.6 | - |
| **PPV(nonTor) %** | 99.8 | 99.8 | 96 | 88 | 99.4 |
| **Overall ACC. %** | **99.1** | **99.8** | **94** | **88.1** | **-** |

## 6 CONCLUSIONS

This paper presents experimental study using two algorithms to detect nonTor traffic in UNB-CIC Tor Network Traffic dataset. The research mainly focuses on detecting nonTor traffic in a representative dataset of real-world traffic to expose the activities within the Tor-traffic that downgrades the privacy of users. The work proposes CFS-ANN hybrid classifier in the detection of nonTor traffic in UNB-CIC Tor Network Traffic dataset. Experimental results show the proposed algorithm detects nonTor with an accuracy of 99.8%, detection rate of 100% and false positive rate of 1.2%. The proposed algorithm performed better than SVM and C4.5 proposed in [7] as shown in table 4. The proposed hybrid classifier reduces the dimensionality of the data size by 65% removing the less effective features thereby lowering computational cost and training time.

In the future, the performance of the proposed algorithm and deep neural networks will be analysed in the classification of the 8 different types of traffic in the UNB-CIC Tor Network Traffic dataset.

## ACKNOWLEDGMENTS

The authors would like to thank the Canadian Institute of Cyber Security, University of Brunswick for providing us with the dataset to carry out this work.